\pdfoutput=1 

\documentclass{WILEYSIX}
\usepackage{amsfonts}
\usepackage{amsmath}
\usepackage{amssymb}
\usepackage{verbatim}
\usepackage{graphicx}
\usepackage{url}

\newcommand{\VT}[1]{\ensuremath{{V_{T#1}}}}

\newbox\sectsavebox
\setbox\sectsavebox=\hbox{\boldmath\VT{xyz}}

\newtheorem{defn}{Definition}
\newtheorem{axm}{Axiom}
\newtheorem{thm}{Theorem}

\newtheorem{cor}{Corollary}

\newtheorem{prf}{Proof}

\def\figs{Figures}

\begin{document}
\chapter{A Methodology for Optimizing Multithreaded System Scalability on Multi-cores}
\chapterauthors{Neil Gunther\affilmark{1},  Shanti Subramanyam\affilmark{2} and Stefan Parvu\affilmark{3}
\chapteraffil{\affilmark{1}Performance Dynamics Company, Castro Valley, California, USA\\
\affilmark{2}Yahoo! Inc., Sunnyvale, California, USA\\
\affilmark{3}Nokia, Espoo, Finland}
}

\offprintinfo{Programming Multi-core and Many-core Computing Systems}{S. Pllana and 
\mbox{F. Xhafa} (eds.)}

\section{Introduction}
The ability to write efficient multithreaded programs is vital for system
scalability, whether it be for parallel scientific codes or large-scale web
applications. Scalability is about guaranteeing sustainable size, so it
should be incorporated into initial system design rather than retrofitted as
an afterthought. That requires a complete methodology which combines
controlled measurements of the multithreaded platform together with a scalability 
modeling framework within which to evaluate those performance measurements.
\index{scalability} \index{performance}

In this chapter we show how scalability can be quantified using the {\em
Universal Scalability Law}~\cite{gcap,usl} by applying it to
controlled performance measurements of memcached, J2EE and Weblogic. 
Commercial multi-core processors are essentially black-boxes and although some
manufacturers do offer specialized registers to measure individual core
utilization~\cite{purr,corestat}, not just overall processor utilization, the
most accessible performance gains are primarily available at the application
level. 
We also demonstrate how our methodology
can identify the most significant performance tuning opportunities to
optimize application scalability, as well as providing an easy means for
exploring other aspects of the multi-core system design space.
\index{Universal scalability law (USL)} \index{methodology} \index{multi-core}

The typical performance focus is on tools and techniques to profile and compile 
fine-grained parallel codes for scientific applications executing on
many-core and multi-core processors. Here, however. we shall be concerned
with performance at the other end of that spectrum, viz., {\em system
performance} of concurrent, multithreaded applications as employed by
commercial enterprises and large-scale web sites. Economies of scale
dictate that these systems eventually be migrated to many-core and multi-core platforms.
\index{multi-core} \index{performance}

Why is the emphasis on system performance important?
Whatever the performance gains attained at the individual processor level,
the impact of those  gains must also be evident at the integrated system
level so as to justify the cost of the effort. A fortiori,
optimizing a local processor subsystem does not guarantee that the total 
system will also be optimized.

The claimed benefits of the various tools used for programming multi-core
applications~\cite{ieee10,masspar,acm11} need to be evaluated {\em quantitatively},
not merely accepted as qualitative prescriptions~\cite{hetero,patterson}. It often
happens that applications which are heralded as being multithreaded and scalable,
turn out not to be when measured correctly~\cite{velocity}. 
To avoid setting incorrect expectations, system performance
analysis should be incorporated into a comprehensive methodology 
rather than being done as an afterthought. We provide such a methodology in
this chapter. \index{methodology} \index{multi-core} \index{performance}

The organization of this chapter is as follows. In Sect.~\ref{sec:mthreads} we
establish some of the terminology used throughout and the basic procedural steps for
assessing system scalability. In Sect.~\ref{sec:measure} we review what it means
to perform the appropriate controlled measurements. The design and
implementation of appropriate load-test workloads for such controlled
measurements is discussed in Sect.~\ref{sec:workload}. Sect.~\ref{sec:uslmodel}
presents the universal scalability model that we use to perform statistical regression on the
performance data obtained controlled measurements. In this way we are able to
quantify scalability. In Sect.~\ref{sec:memcached} we present the first detailed
application of our methodology to quantify memcached scalability. In
Sect.~\ref{sec:multiapps} we give some idea of how to extend our methodology to
a multithreaded java application.
Sect.~\ref{sec:gpus} discusses some ideas about quantifying GPU and many-core
scalability. The importance of our methodology for the often overlooked
validation of complex performance measurements is presented in
Sect.~\ref{sec:validation}. Finally, Sect.~\ref{sec:summary} provides a summary and
possible extensions to our methodology. \index{Universal scalability law (USL)} 
\index{methodology} \index{workload definition} \index{performance}

Although we shall focus on the broader issues of general-purpose, highly- 
concurrent, multithreaded and multi-core~\cite{ieee10} applications~\cite{shanti04},  
we anticipate that readers who are more involved with scientific applications 
will also be able to apply our methodology to their systems. 
\index{scalability} \index{methodology} \index{multi-core}

\section{Multithreading And Scalability} \label{sec:mthreads} 
We begin by presenting the context and terminology for comparing multi-threaded
applications that either scale-out or scale-up. \index{scalability} \index{multi-threading}

Much of the FOSS stack used for running web applications e.g., memcached, MySQL,
Ruby-on-Rails, has scalability limitations that are masked by the widespread
adoption of horizontal scale-out. As traffic growth forces the necessity for more and 
cheaper multi-core servers, multithreading scalability becomes a significant
issue once again. \index{multi-core}

Most web deployments have now standardized on horizontal scale-out in every
tier---web, application, caching and database---using cheap, off-the-shelf,
white boxes. In this approach, there are no real expectations for vertical
scalability of server applications like memcached or the full LAMP stack. But
with the potential for highly concurrent scalability offered by newer multi-core
processors, it is no longer cost-effective to ignore the potential under
utilization of processor resources due to poor thread-level scalability of the
web stack. \index{scalability} \index{multi-core}

Our USL methodology quantifies scalability using the following iterative procedure:
\begin{enumerate}
\item 
Measure the system throughput (e.g., requests per second) for a configuration 
where either the number of user-threads is varied on a fixed multi-core platform or 
the number of physical cores is varied using a fixed number of user-threads per core. 
\item 
Measurements should include at least half a dozen data points in order to make the  
regression analysis statistically meaningful.
\item 
Calculate the capacity ratio $C(N)$ and efficiency $E(N)$ defined in Section \ref{sec:uslmodel}.
\item 
Perform nonlinear statistical regression~\cite{regressR} to determine the USL scalability
parameters $\alpha$, $\beta$ defined in Sect.~\ref{sec:uslmodel}.
\item 
Use the values of $\alpha$ and $\beta$ to predict $N_c$, 
where the scalability maximum is expected to occur. 
$N_c$ may lie outside any physically attainable system configuration.
\item \label{step:synergy}
The magnitude of the $\alpha$ parameter is associated with system {\em contention} effects
(in the application, the hardware or both), and the $\beta$ parameter is
associated with data {\em coherency} effects. This step provides the vital
connection between the numerical output of the USL model and the identification
of likely candidates for further performance tuning in software and hardware.
See Sects.~\ref{sec:memcached} and~\ref{sec:validation}.
\item
Repeat these steps with a new set of measurements until any differences 
between data and the USL projections are optimized.
\end{enumerate}
We elaborate on each of these steps in the subsequent sections.
\index{scalability}

\section{Controlled Performance Measurements} \label{sec:measure}
\begin{figure}[ht]  
\centerline{\includegraphics[scale=0.5]{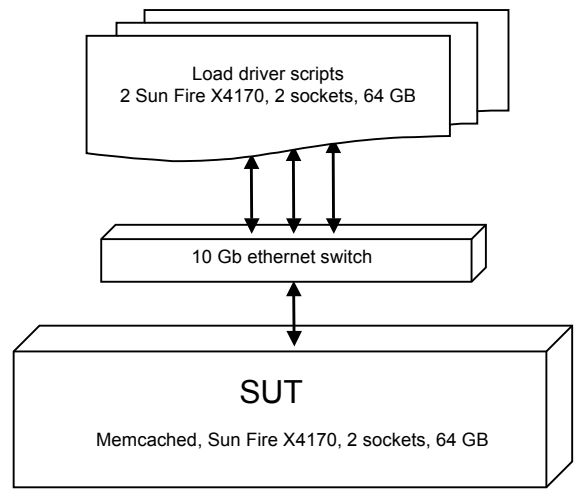}}
\caption{Schematic of scalability load measurement configuration.}  \label{fig:testrig}
\end{figure}

When doing scalability analysis of multithreaded applications, it is important
to collect the data using controlled measurements. Controlled measurements 
require:
\begin{enumerate}
\item
a controlled hardware platform that faithfully represents the real system being
analyzed. The load-test platform that we used to perform the measurements 
presented in Sect.~\ref{sec:memcached} is shown schematically in Fig.~\ref{fig:testrig}.
\item 
a well-designed workload together with tools that produce accurate data
resulting in measurements that are repeatable. The workloads that we used are
described in Sect.~\ref{sec:workload}. \index{workload definition}
\end{enumerate}

\begin{figure}[hb]
\centerline{\includegraphics[scale=0.5]{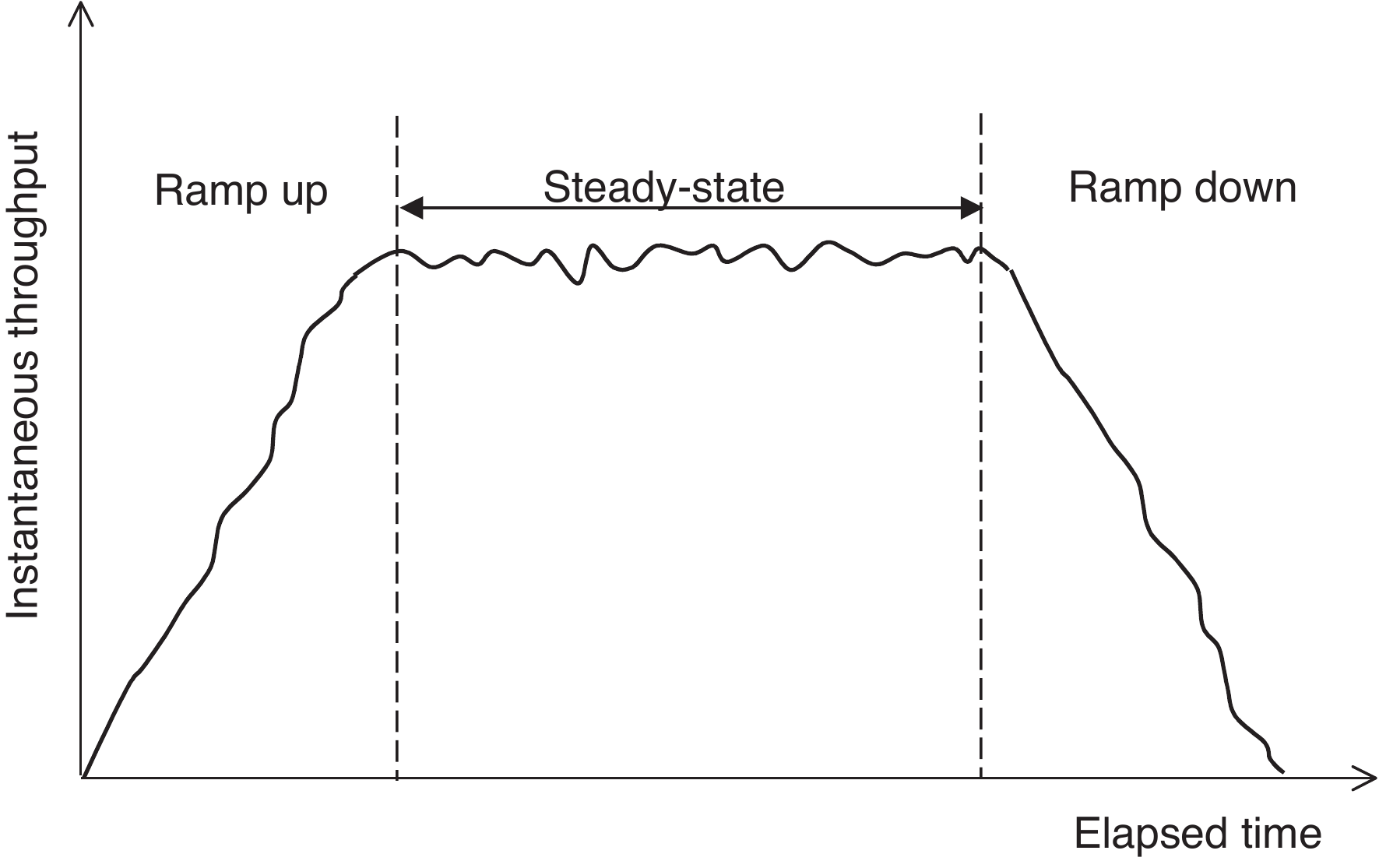}}
\caption{Steady state measurement of instantaneous throughput, for a particular load value $N$,  
represented as a time series to show the ramp up and ramp down phases.}
\label{fig:sstatex}
\end{figure}

The throughput results from a typical performance test are shown in Figure
\ref{fig:sstatex}. A performance test is characterized by a ``ramp-up'' period
in which load is increased on the system, a ``steady-state'' period during which
performance data is gathered and a ``ramp-down'' period as the load diminishes. 
\index{performance}

It is important to ensure that the ramp-up period is sufficiently large to get
the server performing operations in a normal manner, e.g., all data that is
likely to be cached has been read in. This can require times ranging from a
couple of minutes to several tens of minutes, depending on the complexity of the
workload. \index{workload definition}

The steady-state time should be sufficiently long to include all of the activity
that may occur on the system during normal operations (e.g. garbage 
collection, writing to logs at some regular interval, etc.)

Scalability tests should ensure that the infrastructure is well-tuned and 
does not have inherent bottlenecks (e.g. incorrect network routes). This implies
active monitoring of the test infrastructure and analysis of the data to
ascertain it is accurate. Repeating the tests can also help to validate measurements.
\index{scalability}

\section{Workload Design and Implementation} \label{sec:workload}
Data collected from controlled 
performance measurements are only as good as the workload used to run 
the tests. A poorly designed workload can result in irrelevant 
measurements and wrong conclusions~\cite{shanti04}. 
A well-designed workload should have the following characteristics: \index{workload definition}
\begin{description}
\item[Predictability:]
The behavior of the system while running the workload, should be predictable. 
This means that one should be able to determine how the workload processes 
requests and accesses data. This helps in analyzing performance. \index{performance}
\item[Repeatability:]
If the workload is run several times in an identical fashion, it should produce 
results that are statistically identical. Without this repeatability, performance 
analysis becomes difficult. \index{workload definition}
\item[Scalability:] \index{scalability}
A workload should be able to place different levels of load in order to test the
scalability of the target application and infrastructure. A workload that can 
only generate a fixed load or one that scales in an haphazard fashion that does 
not resemble actual scaling in production is not very useful. \index{workload definition}
\end{description}

These characteristics can be realized using the following design principles:
\begin{description}
\item[Design the interactions:] 
Define the actors and their use cases. The use cases help define the 
operations of the workload. In a complex workload, the different actors may 
have different use cases, e.g., a salesperson entering an order and an 
accountant generating a report. Combine use cases that are likely to occur
together into a workload. For example, batch operations run at night should be
a separate workload from online, interactive operations. \index{workload definition}

\item[Define the metrics:] 
Typical metrics include throughput (number of operations executed per unit
time) and response time. Response time metrics are usually specified as
average, 90th or 95th percentiles.

\item[Design the load:] 
This means defining the manner by which the different operations 
associated with the metrics offer load onto the servers. 
This involves deciding on the operation 
mix, mechanisms to generate data for the requests, and deciding on arrival rates
or think times. This step is crucial to get right if the workload needs to 
emulate a production system and/or is being used for performance testing of the
important code paths in the application. A slow operation that is executed only
1\% of the time can sometimes be ignored whereas even a 5\% drop in performance of
an operation that occurs 50\% of the time may be not be tolerable. 
\index{workload definition} \index{performance}

\item[Define scaling rules:] \label{item:step4}
This step is often overlooked, leading to overly optimistic results during testing. 
Complex workloads need a means by which to scale the workload depending on the actual 
deployment hardware. Often, scaling is done by increasing the 
number of emulated users/threads. Any data-dependent application also needs to 
have the data-set scaled in order to truly measure the performance impact of a 
large number of concurrent users. \index{workload definition} \index{performance}
\end{description}

With regard to workload implementation, there exist several open-source and
commercial tools that aid in the development of workloads, and running them.
Available tools vary considerably in functionality, ability to scale, and
their own performance overhead. Some preliminary investigations may be
necessary to ensure that a given choice of tool can meet the anticipated
requirements. \index{workload definition} \index{performance}

\section{Quantifying Scalability} \label{sec:uslmodel}
There are many well known techniques for achieving better scalability: collocation,
caching, pooling, and parallelism, to name a few. But these are only 
qualitative descriptions. How can one decide on the relative merits of any of these 
techniques unless they can be quantified? This is clearly a role for
performance modeling. Performance models are essential, not only for
prediction but, as we discuss in Sects.~\ref{sec:memcached} and \ref{sec:validation},  
for {\em interpreting} scalability measurements. \index{scalability} \index{modeling} 
\index{performance}

Many performance modeling tools, such as event-based simulators and analytic
solvers, are based on a queueing paradigm that requires measured service times
as modeling inputs. More often than not, however, such measurements are
unavailable, thereby thwarting the use of these modeling tools. This is
especially true for multi-tier, web-based applications. A more practical
intermediate approach is to apply nonlinear regression~\cite{regressR} to  performance
measurements that are more accessible; the major advantage being that service-time
measurements are not required. \index{modeling} \index{queueing theory}

\subsection{Queueing Model Foundations} \label{sec:uslqueues}
The universal scalability model (or USL model) that we present
in this section is a realization of the approach alluded to in the previous section.
The USL is a nonlinear parametric
model~\cite{gcap,usl} derived from a well defined queue-theoretic model known as the 
{\em machine repairman model}~\cite{gross,ppdq}.  \index{scalability} 
\index{Universal scalability law (USL)} \index{queueing theory}

Elementary $M/M/m$ queueing models~\cite{ppdq} of multi-cores and multithreaded systems are
too simple because they allow an unbounded number of requests to occupy the
system and they cannot account for processor-to-processor interactions.
Machine-repairman models, like $M/G/m/N/N$, are defined to have only
a finite number of requests~\cite{ppdq}. That constraint can be used to reflect the finite
number of threads in a load test platform, as discussed on
Sects.~\ref{sec:measure} and~\ref{sec:workload}. Alternatively, $M/G/m/N/N$
models can represent the interactions between $N$ processors~\cite{balbo}.
Indeed, the machine repairman model can be further generalized in terms of
queueing network models to analyze the performance of parallel
systems~\cite{sevcik}, including architectures with multiple latency
stages~\cite{pdcs05}, provided the requisite service times can be measured.
\index{queueing theory} \index{multi-core} \index{performance}

\begin{table}
\centering
\caption{Interpretation of $M/M/1/N/N$ queueing metrics} \label{tab:qmets}
\begin{tabular}{clll}	
\hline
Metric 	& Repairman 		& Multi-core 			& Multi-thread\\
\hline
N 		& machines 			& virtual processors			& user threads\\
Z 		& up time 			& execution period		& think time\\
S 		& service time		& transmission time 	& processing time\\
W	 	& wait time 		& interconnect latency	& scheduling time\\
X		& failure rate    	& bus bandwidth				& throughput\\
\hline
\end{tabular}
\end{table}

Here, we restrict ourselves to the $M/M/1/N/N$ queueing model where the single
Markovian server represents the interconnect latency between $N$ processors or
cores. Since the components of this queue have a consistent physical
interpretation with respect to multi-core performance metrics (Table~\ref{tab:qmets}),
we also avoid mere curve fitting exercises with ad hoc parameters. \index{queueing theory}
\index{multi-core} \index{Markov process}

\begin{figure}[ht]
\centerline{\includegraphics[scale=1.0]{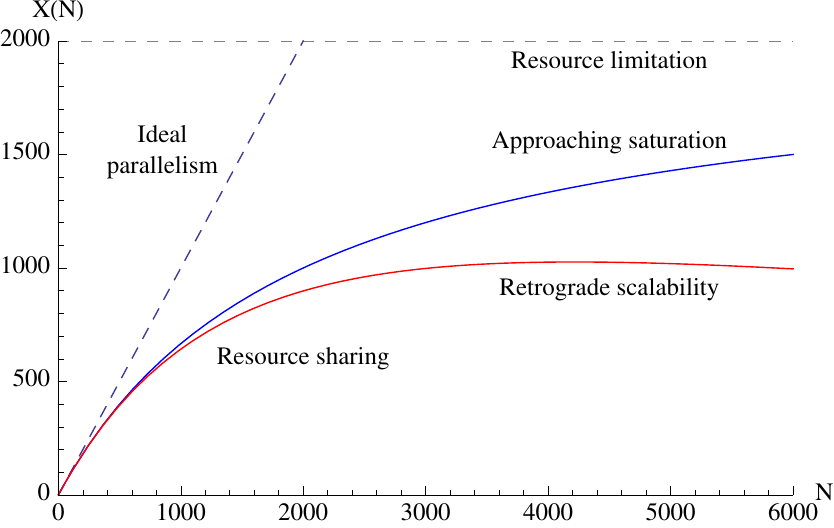}}
\caption{Physical components of scalability.} \label{fig:usl-parts}
\end{figure}

To motivate our choice of performance model, we briefly review 
the key physical attributes of scalability. Referring to Fig.~\ref{fig:usl-parts}:
\begin{enumerate}
\item {\bf Ideal parallelism:} 
Linear scaling corresponds to  {\em equal bang for the buck}
computational capacity where each increment in the load, $\Delta N$ on
the $x$-axis, produces a constant increment in throughput, $\Delta X$ on the
$y$-axis, as indicated by the dashed inclined line in
Fig.~\ref{fig:usl-parts}. Such linearity in capacity can be written
symbolically as:
\begin{equation}
C(N) = N
\end{equation}
This includes scaled sizing of the workload~\cite{gcap}.
\index{workload definition}

\item {\bf Resource sharing:} 
Accounts for the fall away from linear scaling due to waiting for access to shared resources.
This loss of linearity due to resource contention is associated with the USL model parameter 
\begin{equation}
0 < \alpha < 1    \label{eqn:seriality}
\end{equation}

\item {\bf Resource limitation:} 
Even if such linear scaling is achievable, it cannot exceed the finite
capacity of the system resources. This is defined by an asymptotic bound from
above:
\begin{equation}
\lim_{N\rightarrow\infty} C(N, \alpha) = \frac{1}{\alpha}
\end{equation}
This saturation limit is shown as the dashed horizontal line in Fig.~\ref{fig:usl-parts}.
This bound could be lower due to execution-time skew in components of 
the workload~\cite{ppa,cilk}. \index{workload definition}

\item {\bf Retrograde scaling:} Worse than saturation, this effect arises from the additional latency 
due to {\em pairwise} interprocessor communication, e.g., exchange of data between caches, 
and is given by the binomial coefficient:
\begin{equation}
\binom{N}{2} = \dfrac{N(N-1)}{2} \label{eqn:binom}
\end{equation}
and is associated with a USL parameter $\beta$.
\end{enumerate}

\noindent
Another useful metric is the efficiency:
\begin{equation}
E(N) =  \dfrac{C(N)}{N}  \label{eqn:efficiency}
\end{equation}
which defines the scalability (\ref{eqn:caprat}) per core or per thread. 
We shall apply this metric to data validation in Sect.~\ref{sec:validation}.
\index{scalability}

\subsection{Universal Scalability Model} \label{sec:usl}
The following theorem allows us to combine each of the physical scalability
components of Sect.~\ref{sec:uslqueues} into a parametric model.
\index{scalability} \index{Universal scalability law (USL)}

\begin{defn}[Universal Scalability Law]
\begin{equation}
C(N, \alpha, \beta) = \dfrac{N}{1 + \alpha (N - 1) + \beta N (N - 1)} \label{eqn:usl}
\end{equation}
where the factor of 2 in (\ref{eqn:binom}) has been absorbed into the $\beta$ coefficient.
\end{defn}

\begin{thm}[Queueing Bound] \label{thm:usl} 
The USL model (\ref{eqn:usl}) is equivalent to the synchronous bound on 
the throughput of a machine-repairman queueing model with a 
service time that is linearly load-dependent on $N$.
\end{thm}

The formal proof is too long to reproduce here. A pivotal observation~\cite{gcap} is that for
$M/M/1/N/N$, the throughput $X(N)$ is bounded below by:
\begin{equation}
X(N) \geq \dfrac{N}{N S + Z}
\end{equation}
in the notation of Table~\ref{tab:qmets}.
It also excludes super-linear scaling~\cite{sevcik,cilk}.

\begin{prf}[Sketch]
When the first request is in 
service (at the repairman) the mean waiting time for the remaining requests is 
\begin{equation}
W = (N-1) S			\label{eqn:sync}
\end{equation} 
where $N$ is the number of requests in the system. 
Let the service time be linearly load-dependent:
\begin{equation*}
S(N) = c N S
\end{equation*}
with $c$ a constant of proportionality. For synchronous queueing 
all requests are enqueued simultaneously, so we can rewrite (\ref{eqn:sync}) as:
\begin{equation}
W = c N (N-1)  \, S \label{eqn:quadp}
\end{equation} 
Expressed as relative relative throughput, (\ref{eqn:quadp}) appears in the
denominator of (\ref{eqn:usl}) as the corresponding quadratic term.  \qed
\end{prf}
The interested reader is referred to Ref.~\cite{usl} for the details and
Ref.~\cite{zones} for supporting simulation results. \index{queueing theory}

Perhaps the most important point for our methodology is that (\ref{eqn:usl}) 
is a mean value equation in the queueing variables and, in that sense, accounts 
for the possibility of fluctuations in the size of workload components and 
sub-tasks. In particular, the machine repairman model has been proven to be robust 
to fluctuations in these queue variables~\cite{gross}.  \index{methodology} 
\index{queueing theory}

\begin{cor}[Duality] \label{cor:duality}
The scaling variable $N$ in the parametric model (\ref{eqn:usl})
can be interpreted equally as representing a finite number of
threads (software view) or a finite number of core processors (hardware
view) because it is a bound on same the $M/M/1/N/N$ queueing model.  \index{queueing theory}
\end{cor}

Setting $\beta=0$ in (\ref{eqn:usl}) produces the
standard parametric version of Amdahl's law with (\ref{eqn:seriality}) the {\em
serial fraction} of the workload. However, by virtue of Theorem~\ref{thm:usl},
Amdahl's law can be interpreted as a limiting case (zero coherency delays) of the
USL. \index{workload definition} \index{Amdahl's law}

\begin{cor}[Amdahl's Law] \label{cor:amdhal} 
Amdahl's law corresponds to the relative throughput (speedup) 
due to synchronous queueing in the standard machine repairman queueing model with 
constant mean service time. \index{queueing theory} \index{Amdahl's law}
\end{cor}

Amdahl's law is the synchronous throughput bound on 
an $M/M/1//N$ queue having a load-independent mean service time S. \index{Amdahl's law}
\begin{prf}[Sketch]
The proof relies on the identity:
\begin{equation}
\alpha = \dfrac{S}{S + Z} \rightarrow  
\begin{cases} 
0	& \text{as}~S \rightarrow 0, ~\text{with}~Z = \text{const.},\\ 
1	& \text{as}~Z \rightarrow 0, ~\text{with}~S = \text{const.}. 
\end{cases} \label{eqn:ident}
\end{equation}
between the queueing metrics in Table~\ref{tab:qmets} and the parameter $\alpha$ in 
(\ref{eqn:seriality}). \qed
\end{prf}
See Appendix A of  Ref.~\cite{gcap} and Ref.~\cite{usl} for a more detailed discussion. 

An important point to note from the preceding is that Amdahl's law represents worst case 
queueing effects~\cite{ieee08,isca}. This is consistent with the notion that synchronous requests have 
longer delays than asynchronous requests.
The latter being the mean value throughput for $M/M/1//N$.
Other examples of applying (\ref{eqn:usl}) to both hardware and software scalability,  
can be found in~\cite{gcap}. \index{scalability} \index{queueing theory} \index{Amdahl's law}

The capacity ratio for measured data is defined as the normalization:
\begin{equation}
C(N) = \frac{X(N)}{X(1)}  \label{eqn:caprat}
\end{equation}
Since the capacity ratio has two definitions---one empirical
(\ref{eqn:caprat}) and the other analytical (\ref{eqn:usl})---the optimization goal is to
match them in such a way that the adjusted USL coefficients 
provide the best fit the performance data.

The key distinction is that, unlike Amdahl's law, (\ref{eqn:usl}) 
possesses a maximum at \index{Amdahl's law}
\begin{equation}
N_c = \sqrt{\frac{1 - \alpha}{\beta}}  \label{eqn:nmax}
\end{equation}
the location of which is controlled by the USL coefficients according to:
\begin{enumerate}
\renewcommand{\labelenumi}{(\alph{enumi})}
\item $N_c \rightarrow 0$ 				as $\alpha \rightarrow 1$ \label{item:sigone}
\item $N_c \rightarrow 0$ as $\beta \rightarrow \infty$ \label{item:kinf} 
\item $N_c \rightarrow \infty$ 			as $\beta \rightarrow 0$ \label{item:kamd} 
\item $N_c \rightarrow \beta^{-1/2}$ 	as $\alpha \rightarrow 0$ \label{item:khalf} 
\end{enumerate}

\noindent
The important implication for our methodology is that beyond $N_c$ the throughput becomes {\em
retrograde}. See Fig.~\ref{fig:usl-parts}. This effect is commonly observed in applications that
involve shared-writable data.  \index{methodology}

Summarizing the steps for application optimization:
\begin{enumerate}
\item Steady state measurements of throughput $X(N)$ for each load point $N$.
\item At least half a dozen $N$ values are required in order to be statistically significant for USL fitting.
\item Calculate the capacity ratio (\ref{eqn:caprat}) for each $N$ value.
\item Use nonlinear statistical regression~\cite{regressR} to determine the USL coefficients $\alpha$ and $\beta$.
\item Optimize the complete scalability function (\ref{eqn:usl}) for any desired $N$ value.
\end{enumerate}
The same methodological procedure can be applied to hardware scalability optimization although,
as we pointed out in the introduction, most commodity hardware is now a silicon black-box 
which mean the hardware performance tuning opportunities are far fewer. 
\index{scalability} \index{performance}

The use of the term ``universal'' in this context refers not only to the general
applicability of (\ref{eqn:usl}) to both multi-core hardware and multi-threaded
software scalability, but also to the fact that no more than two coefficients
are needed to accommodate the possibility of reaching saturation limits (Amdahl
scaling) or thrashing limits (coherency delays). in the latter case, there is little virtue
on modeling such degraded performance; better to try and improve it. \index{scalability}
\index{modeling} \index{multi-core} \index{multi-threading} \index{performance}

We now present some case studies that demonstrate how this methodology has been 
successfully applied. \index{methodology}

\section{Case Study: Memcached Scalability} \label{sec:memcached}
As mentioned in Sect.~\ref{sec:mthreads},
most large-scale web sites have standardized on horizontal scale-out in every tier
as a simple way to achieve high degrees of scalability.
A ubiquitous application used in this context is {\em memcached} (MCD).
In this section, 
we demonstrate how are our analysis leads to improved thread scalability of 
MCD~\cite{velocity}. \index{scalability}

\begin{figure}[ht]
\sidebyside{
\centerline{\includegraphics[scale=0.675]{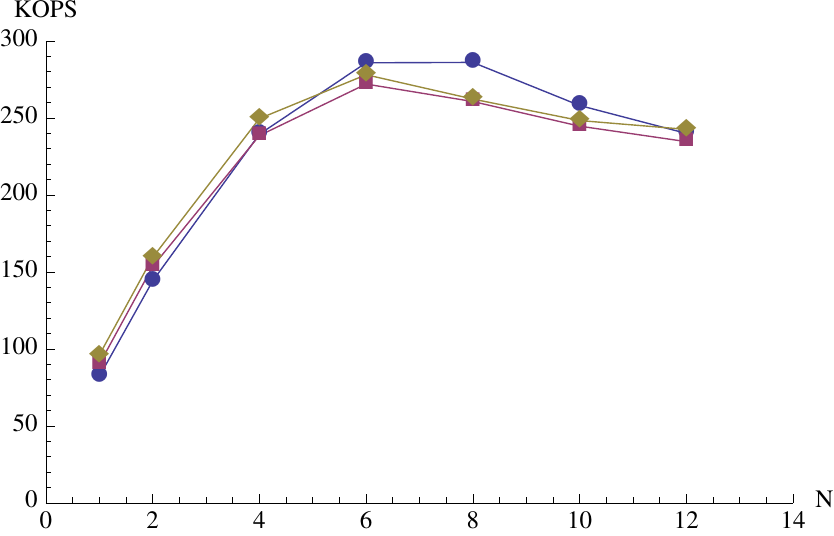}}
\caption{Throughput of three MCD releases up $N=12$ threads.}
\label{fig:mcd-data3}
}
{
\centerline{\includegraphics[scale=0.675]{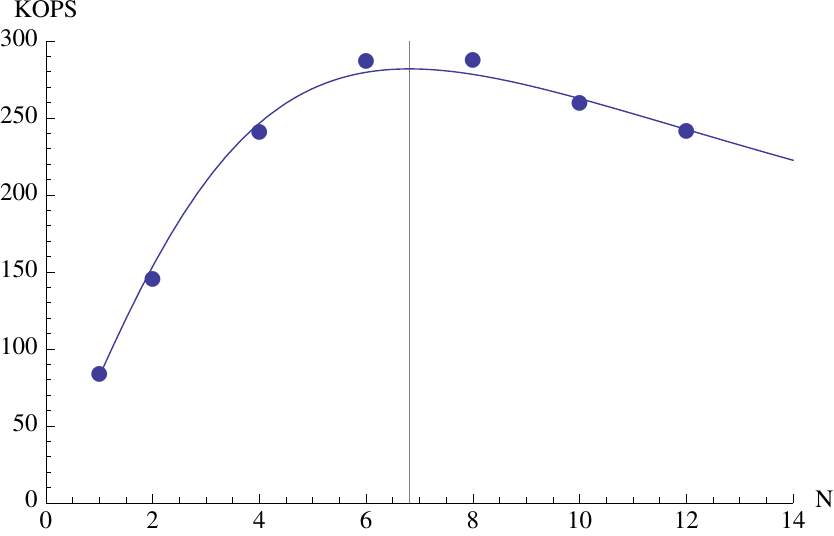}}
\caption{USL regression analysis of memcached 1.2.8 data ({\em dots}) in
Fig.~\ref{fig:mcd-data3}.}
\label{fig:mcd128-usl}
}
\end{figure}

Figure~\ref{fig:mcd-data3} shows controlled 
MCD throughput  as a function of $N \leq 12$ threads for three release: 1.2.8,
1.4.1 and 1.4.5, measured in thousands of operations/sec (KOPS). Each release has
a very similar retrograde throughput profile, peaking between $N=6$ and $N=7$ threads.

Whereas the lines in Fig.~\ref{fig:mcd-data3} merely associate data points
belonging to the same MCD release, the curve in Fig.~\ref{fig:mcd128-usl} is
generated by statistically fitting (\ref{eqn:usl}) to those data and is 
not required to pass through every data point.

The USL regression analysis of MCD 1.2.8 reveals a contention parameter value of 
$\alpha = 0.0255$ and a coherency parameter value of $\beta = 0.0210$.
Repeating this procedure with the other MCD versions results in the USL 
coefficients summarized in Table~\ref{tab:mcdparams}. In this way, the 
scalability of MCD is now fully quantified. It is also clear that it would be desirable 
to move the estimated maximum at $N_c \approx 6$ to a higher value. \index{scalability}

\begin{table}[htdp]
\caption{Memcached scalability parameters} \label{tab:mcdparams}
\begin{center}
\begin{tabular}{cccc}
\hline
Version 		& $\alpha$ 	& $\beta$ 	& $N_c$\\
\hline
1.2.8 	& 0.0255 	& 0.0210		& 6.8121\\
1.4.1 	& 0.0821 	& 0.0207		& 6.6591\\
1.4.5 	& 0.0988 	& 0.0209		& 6.5666\\
\hline
\end{tabular}
\end{center}
\label{tab:mcd}
\end{table}

\begin{figure}[ht]
\sidebyside{
\centerline{\includegraphics[scale=0.675]{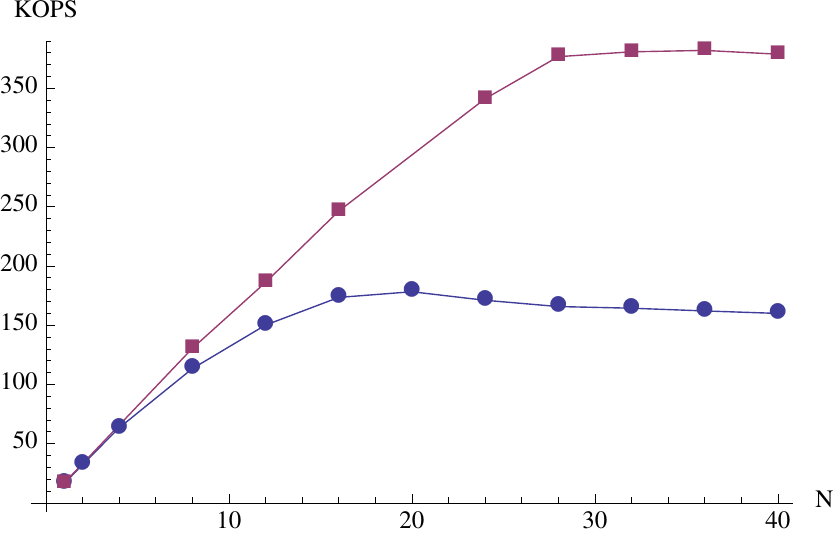}}
\caption{Comparison of scalability data for standard MCD ({\em lower curve}) 
and patched MCD ({\em upper curve}).} \label{fig:patch-data}
}
{
\centerline{\includegraphics[scale=0.675]{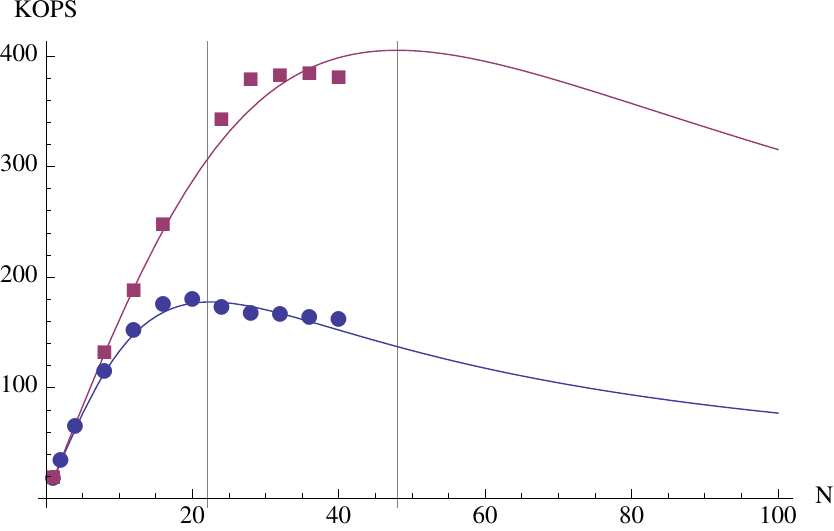}}
\caption{USL model of patched MCD data in Fig.~\ref{fig:patch-data} extended out
to $N=100$ threads.}  \label{fig:patch-usl}
}
\end{figure}

Fig.~\ref{fig:patch-data} shows how scalability improved after various code changes 
were applied. This is where the procedural steps of our USL methodology, outlined 
in Sect.~\ref{sec:mthreads}, actually pay off. \index{scalability} \index{methodology}

Access to the MCD cache is controlled by a single mutex lock. When running
with greater than $6$ threads, contention for this mutex increases dramatically.
A partitioned cache was implemented with each partition controlled by its own
mutex. In addition, contention for the stats lock was identified. This lock
controls access to the stats structure that is updated on every request. The
stats structure was redesigned to hold stats on a per-thread basis. This fix was
applied in release 1.4.5 and they greatly improved scalability. \index{scalability}

The USL fit to the patched MCD data in Fig.~\ref{fig:patch-data} is extended out
to $N=100$ threads in Fig.~\ref{fig:patch-usl}. The original scalability peak at
$N_c=22$ threads ({\em lower curve}) is now moved out to $N_c=48$ threads ({\em
upper curve}). \index{scalability}

At this point, the reader may be dumbstruck as to how the actual changes made to
the application code are determined from the seemingly abstract numerical output
of the USL model. First, it is important to recall that the role of the
performance analyst is to measure and validate, not to modify hardware or
software  for which he or she was not responsible in the first place. That is
the role of the hardware engineer or the software developer. \index{performance}

Second, the interpretation of the USL analysis and the choice of performance tuning 
optimization arises from discussions between the performance analyst and the 
appropriate engineers. Since the latter are the real experts, it is helpful if the 
modeling analysis can point to specific types of effects that may be contributing to inferior 
scalability. This is precisely what the USL does by virtue of its parameters 
having explicit physical meaning, viz., the respective degrees of 
concurrency ($N$), contention ($\alpha$), and coherency ($\beta$). \index{scalability} 
\index{modeling}

In this way, step~\ref{step:synergy} of the USL methodology in Sect.~\ref{sec:mthreads}
can evoke a ``light bulb'' moment for engineers.
In practice, we have seen this synergy occurring time and again. Moreover, the corrective 
action taken is usually something we, as performance analysts, could never have foreseen 
because we were not in possession of the implementation details. Although we have 
presented an example of improvements made to MCD software, 
Corollary~\ref{cor:duality} implies it could also have been that 
scalability improvements came from hardware changes, such as memory resizing or 
more recent revisions to the multicore architecture.
That said, no matter what insights are favored or what tuning actions are adopted, 
the ultimate arbiter is the next iteration of the USL methodology.
\index{scalability} \index{methodology} \index{performance}

\section{Other Multi-threaded Applications} \label{sec:multiapps}
We focused on memcached scalability in Sect.~\ref{sec:memcached} to demonstrate
how the USL methodology is applied in detail. In this section, we show how the
same methodology can be applied to other multi-threaded applications. \index{scalability} 
\index{multi-threading}

Java Enterprise Edition (J2EE) applications are extremely popular in enterprises
because the J2EE platform is known to be robust, secure and
scalable~\cite{shanti05}.

\begin{figure}[ht]
\sidebyside{
\centerline{\includegraphics[scale=0.675]{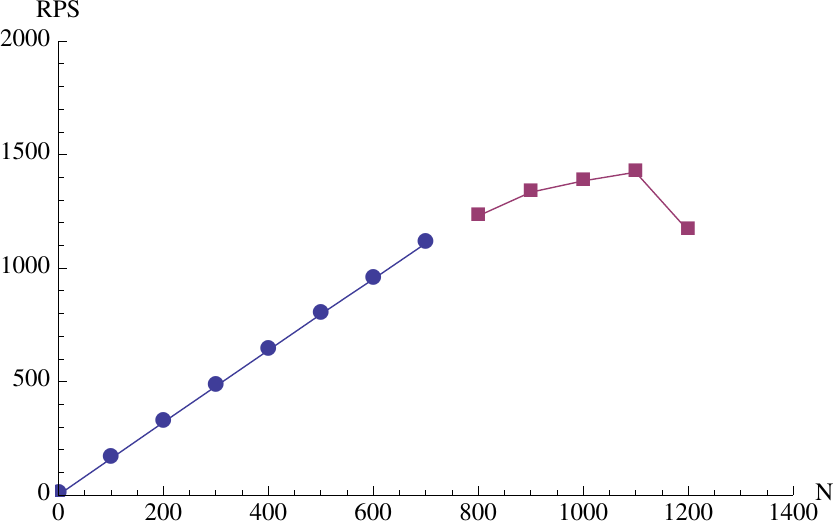}}
\caption{Initial J2EE throughput data ({\em left}) and subsequent data ({\em right})
measured in requests per second (RPS).} 
\label{fig:japp-data}
}
{
\centerline{\includegraphics[scale=0.675]{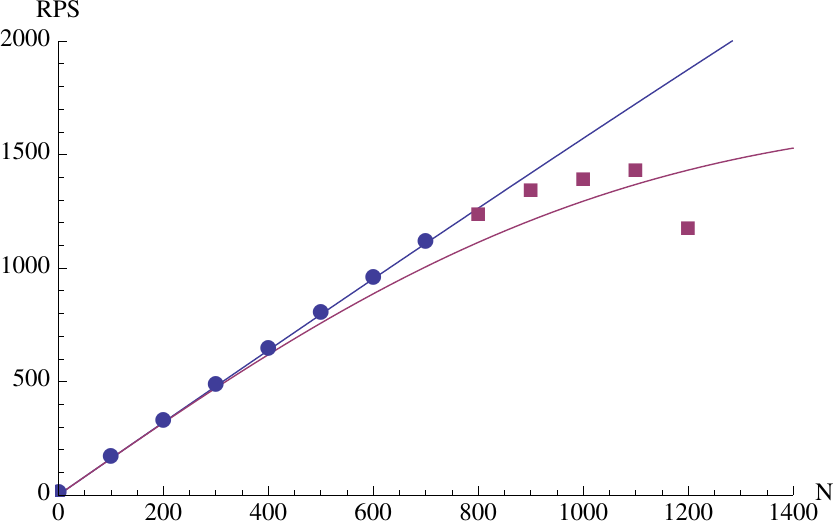}}
\caption{USL models of J2EE scalability for initial data ({\em uppsr curve})
and subsequent data ({\em lower curve}).} \label{fig:japp-usl}
}
\end{figure}

Figure~\ref{fig:japp-data} shows how the application  
throughput scales when load is added to a J2EE server.
The blue and red points show different data sets to exhibit how the 
USL regression values change accordingly. See Table~\ref{tab:j2eeparams}.

\begin{table}[htdp]
\caption{J2EE application scalability parameters} \label{tab:j2eeparams}
\begin{center}
\begin{tabular}{cccc}
\hline
Load 	& $\alpha$ 	& $\beta$ 	& $N_c$\\
\hline
Low 	& $1.49 \times 10^{-5}$ & $6.7 \times 10^{-9}$	& 12216.90\\
High 	& 0.0 					& $2.4 \times10^{-7}$	&  2041.24\\
\hline
\end{tabular}
\end{center}
\label{tab:java}
\end{table}

When the USL is applied to the blue data points, it results in the upper curve
in Fig.~\ref{fig:japp-usl}, which indicates excellent scalability up to more
than \mbox{$N_c \approx 12,000$} users. This modeling result is reasonable as
the initial data set shows almost linear scalability through $N=800$ users.
\index{scalability} \index{modeling}

However, consider what happens when the data set changes. As the load was
increased and the red data points in Fig.~\ref{fig:japp-data} were measured, the
corresponding USL parameters change accordingly resulting in the lower curve of  
Fig.~\ref{fig:japp-usl}. 
The larger $\beta$ value in Table~\ref{tab:j2eeparams} reflects a substantial
decrease in predicted scalability. However, even that scalability is still extremely good
(cf. the corresponding $\alpha$, and $\beta$ for MCD  in
Table~\ref{tab:mcdparams}), but instead of simply appealing to qualitative
descriptions like ``highly scalable,'' or ``great performance,'' the USL
coefficients provides us with true quantification of J2EE scalability.
\index{scalability} \index{performance}

We want to underscore that what  looks like a bad prediction is, in fact,  
precisely how our methodology should work. Based on the initial data,
maximum scalability 
was estimated to occur at $N_c \approx 12,000$ user threads.
Further measurements, however, show that this maximum  occurs at 
$N_c \approx 2,000$ user threads instead. It is not that the original USL projections were 
wrong, but that those initial data did not contain any information about a subsequent 
scaling limitation present in the JVM. 
The important point is that the USL sets expectations and then forces performance engineers to 
explain subsequent deviations at each stage of the measurement process.
\index{scalability} \index{methodology} \index{performance}

\section{Case Study: Data Validation} \label{sec:validation}  
Another simple and immediate practical benefit of applying the methodology 
is {\em validation} of performance data.

\begin{figure}[ht]
\centerline{\includegraphics[scale=0.675]{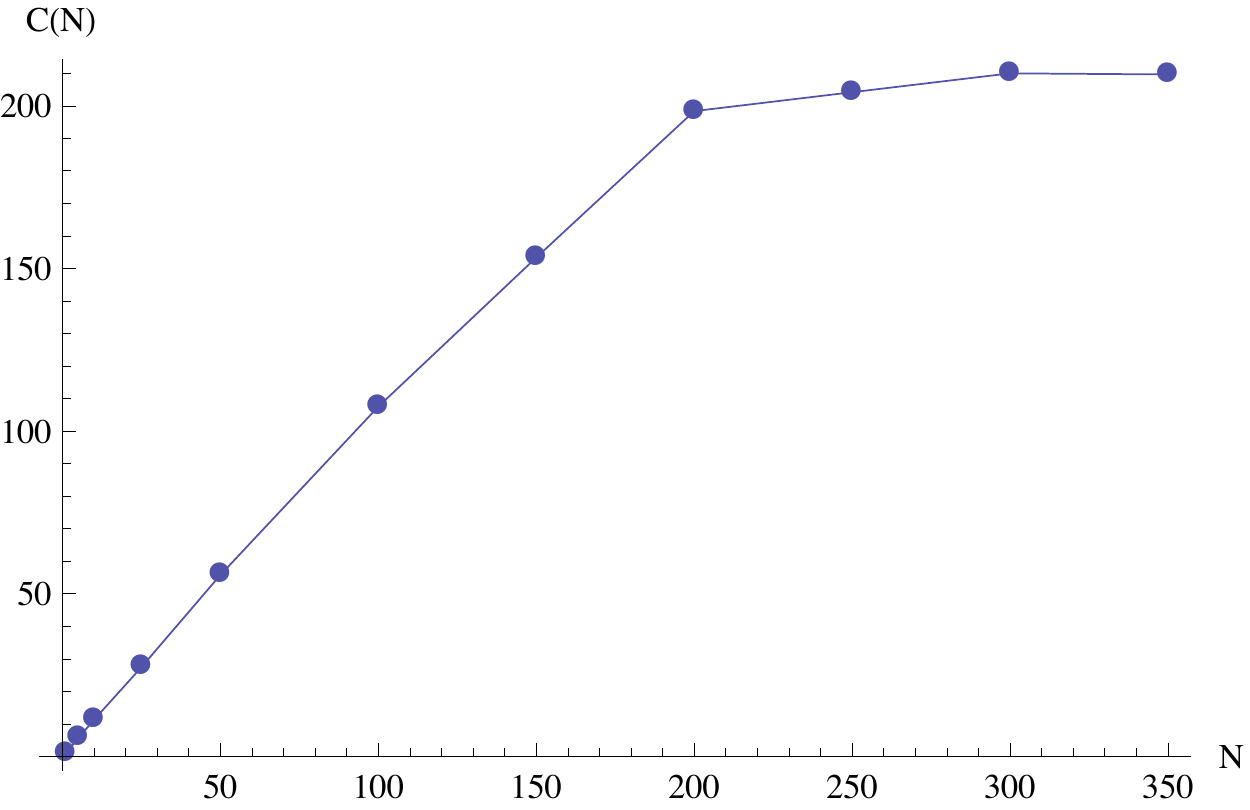}}
\caption{Web application scalability data.} \label{fig:mobi-plot}
\end{figure}

Consider a test environment, similar to that described in
Sects.~\ref{sec:workload} and~\ref{sec:multiapps}, where test scripts are
developed using different test cases for a particular application. In this
case, the test configuration used Apache Jakarta JMeter as a load injector
for a J2EE application~\cite{shanti05} running on Java 6 with a Weblogic
application server. We are not interested in what
this application was doing internally but rather, to examine and validate the load
testing procedure. \index{workload definition}

Several tests were run against this J2EE application and the reported JMeter
values were recorded. As usual, we were interested in the throughput and response time metrics.
Following Sect.~\ref{sec:usl}, throughout was the primary metric of interest for determining 
application scalability. Fig.~\ref{fig:mobi-plot} shows an example of the 
throughput data. It appears acceptable because: \index{scalability}
\begin{itemize}
\item 
The data points are monotonically increasing. A sequence of numbers is
monotonically increasing if each element in the sequence is larger than its
predecessor. Notice  that the profile appears to decrease slightly beyond 
$N = 300$. The USL is designed to model such a characteristic.
\item 
The sequence is linear-rising up to $N = 200$ virtual users.
\item 
The throughput reaches saturation around $N = 300$. This is exactly what we
expect for a closed queueing system~\cite{ppdq} with a finite number of active requests (as is true
for any load-testing or benchmarking system). In this case, the onset of
saturation looks rather sudden as indicated by the discontinuity in the gradient
(``sharp knee'') at $N = 200$. This is usually a sign of significant internal 
 change in the dynamics of the combined hardware-software system.
\end{itemize}

These data seem to pass the visualization test and most performance testing  
would stop here. Unfortunately, visualization alone is not always sufficient proof of 
optimal scalability.
Applying our USL methodology, we modeled these data to 
evaluate the $\alpha$ and $\beta$ coefficients and determine if application 
scalability could further be improved. \index{scalability} \index{methodology} \index{performance}

In setting up the USL model to perform statistical regression, we 
detected some efficiencies (\ref{eqn:efficiency}) that were greater than 100\%. 
In particular, Table~\ref{tab:badmobi} exhibits 
$E(N) > 1$ for test loads in the range $N=5$ to $150$ virtual users.

\begin{table}[htdp]
\caption{Preparation for USL analysis of the data in
Fig.~\ref{fig:mobi-plot}} \label{tab:badmobi}
\begin{center}
\begin{tabular}{rrr}
\hline
$N$ 	& $C(N)$	& $E(N)$\\
\hline
 1 & 1.00 & 1.00 \\
 5 & 5.67 & 1.13 \\
 10 & 11.33 & 1.13 \\
 25 & 27.50 & 1.10 \\
 50 & 55.83 & 1.12 \\
 100 & 107.50 & 1.08 \\
 150 & 153.33 & 1.02 \\
 200 & 198.33 & 0.99 \\
 250 & 204.17 & 0.82 \\
 300 & 210.00 & 0.70 \\
 350 & 209.67 & 0.60\\
\hline
\end{tabular}
\end{center}
\end{table}

From a logical standpoint, we cannot have more than $100\%$ of anything.
Sometimes, however, there are {\em conventions} in performance analysis
where quantities exceeding $100\%$ have a particular interpretation, e.g.,
3200\% processor capacity might be shorthand for a maximal machine utilization of $32$
cores running at $100\%$ busy. Conventions notwithstanding, any numbers that are
out of bounds should be flagged for explanation by performance 
engineers or application developers.

\begin{axm} \label{ax:insight}
\text{Data} + \text{Models} = \text{Insight}  \label{eqn:insight}
\end{axm}

All measurements contains errors and the more complex
the measurement system, the more prone it will be to generating erroneous performance data. 
Without a validation framework, how can it be known when the  
data are wrong? The USL provides a simple mathematical reference 
framework for detecting anomalies like those in Table~\ref{tab:badmobi}. 
We encapsulate this observation in Axiom~\ref{ax:insight}~\footnote{A hybridization 
of the book title {\em Algorithms + Data Structures = Programs} by N. Wirth 
and R. Hamming's observation that computing is about insight, not numbers.}.

So, what was causing the excessive efficiencies in this case? Since we had not
even invoked statistical regression at that point, we knew that the culprit
could not be the USL model. Instead, it became clear that something was amiss
with the measurement process. (Not the usual conclusion) The performance
engineers then set about eliminating one factor at a time. Eventually, it
emerged that the JMeter tool itself was the only remaining explanation for the
source of the erroneous measurements. Without being forced by the USL modeling
framework to resolve this unforeseen issue, further load testing would have been a
waste of time and resources. \index{modeling}

\section{Scalability on Many-core Architectures} \label{sec:gpus}
Tools for writing applications for CPU-GPU many-cores are constantly improving.
Measuring and quantifying many-cores scalability of such applications is the
next step and that requires a methodology, not just tools. In this section we
indicate how the USL methodology can be applied to workloads running
on many-core architectures. \index{scalability} \index{methodology}
\index{workload definition}

\subsection{Trends in Multiprocessing, Multi-cores and Many-cores}
In recent years, vendors have been considering multi-core architectures and how
applications can be migrated from single processors to multiple
processors. In this paradigm, the multi-core forces the
application programmer to focus on maintaining and maximizing execution speed of
a sequential workload but replicating it across multiple processing units inside
the same physical processor~\cite{masspar}. \index{multi-core} \index{workload definition}

A different approach, using many-cores, focuses on how to maximize the aggregate  
throughput; an essential requirement for the gaming industry and anything else 
involving 3D graphics. This many-core approach deploys a much higher number of cores
per physical processor unit, without the need for
internal cache memories, 
logic control unit for executing instructions, 
and other complexities associated with multicore processors.

These alternative paradigms allow developers to consider which is the best
option for their applications. Recent improvements offer additional mechanisms
to select and direct parts of the application to either CPU or GPU, depending of
its intended usage. Compute-intensive sections can be dynamically 
directed to a many-cores processor, 
while single-threaded sections can be assigned to a multicore processor. Such
combinations of CPU and GPU, let the workloads run optimally by taking
intelligent advantage of the type of processor hardware available. \index{workload definition}

However, not all applications are written to take advantage of these new architectures.
For example, legacy single-threaded workloads typically cannot make use of these new options. 
When executed on many-core processors, such workloads will underperform. \index{workload definition}

Without significant modification and porting effort, legacy workloads  
cannot scale well. Testing and analyzing these workloads in a controlled 
fashion (see Sect.~\ref{sec:measure}) is a necessity and presents another 
opportunity for our USL methodology. \index{methodology} \index{workload definition}

\subsection{USL Methodology for GPUs}
Since the USL methodology is generic, it should be applicable to
quantifying the scalability of many-core applications.
In this vein, we have applied it to data kindly provided to us by Prof. Frank Dehne and Kumanan Yogaratnam 
at Carleton University~\cite{cangpu}. \index{scalability} \index{methodology}

\begin{figure}[ht]
\centerline{\includegraphics[scale=0.5]{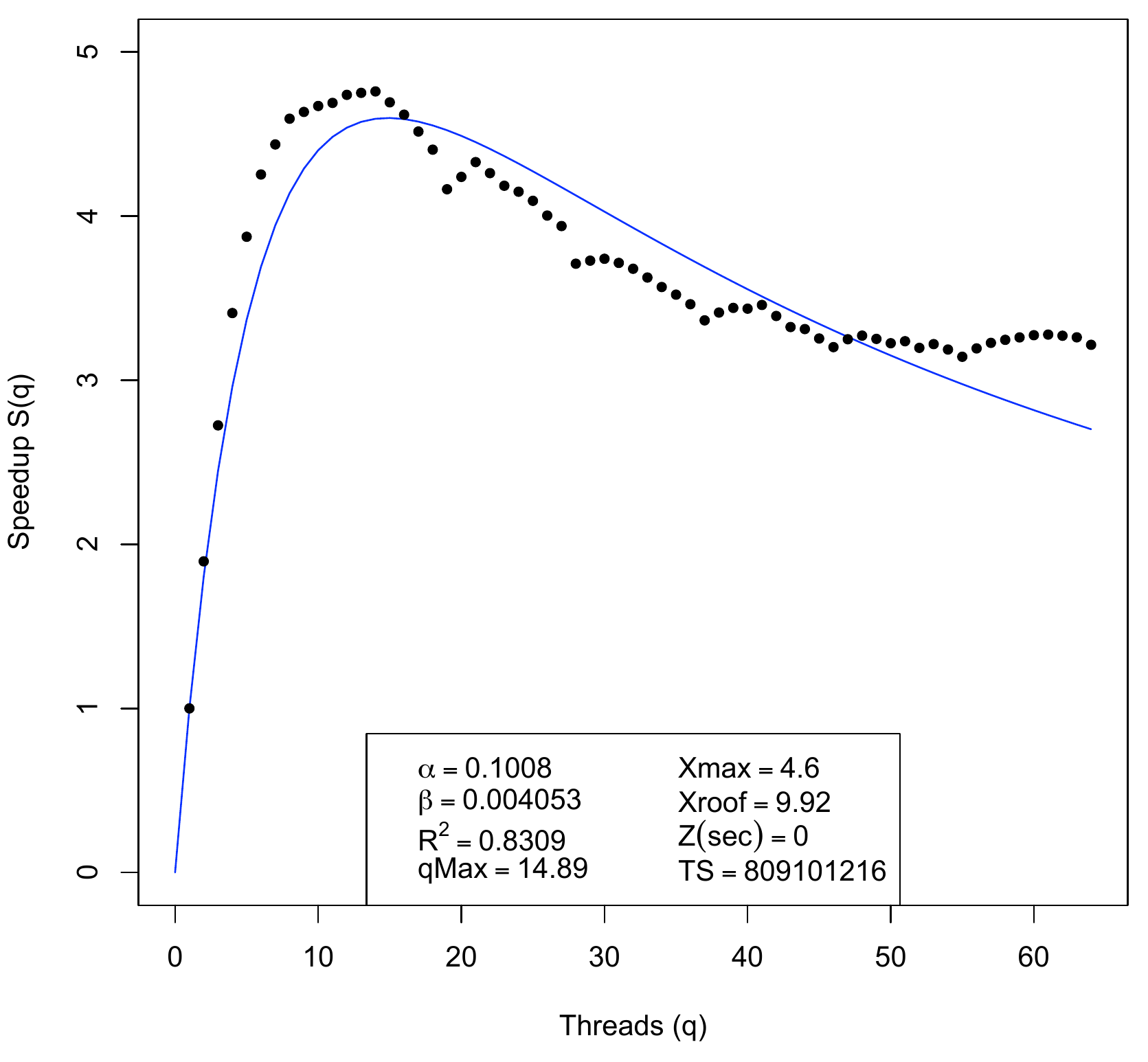}}
\caption{USL fit to nVIDIA Tesla GPU data.} \label{fig:uslgpu}
\end{figure}

They compare the speedup of different parallel graph algorithms running on an 
nVIDIA GeForce 260 with 216 2.1 GHz GPU-cores and 896 MB of RAM.
The parallel speedup is logically equivalent to $C(N, \alpha, \beta)$ in the 
USL formalism.

Their choice of parallel graphing algorithms reveal irregular data access
patterns (shown as dots in Fig.~\ref{fig:uslgpu}) that are different from than
regular data access patterns found in typical parallel processing workloads for
image processing, linear algebra or scientific computing. More importantly, 
significant speedup degradation is observed for $N > 15$ threads.
\index{workload definition}

\section{Future Work}
Applying USL regression analysis to the data in Sect.~\ref{sec:gpus} produces 
the curve in Fig.~\ref{fig:uslgpu}, which has a contention
parameter value of $\alpha = 0.1008$ and a coherency parameter value of $\beta =
0.00405$, with an estimated maximum at $N_c=14.89$ threads.
The interpretation of these coefficients is still under investigation for 
potential ways to improve GPU scalability. In the meantime, the important point
is that these controlled measurements are being compared with the USL performance 
model, thereby reinforcing Axiom~\ref{ax:insight}. \index{scalability} \index{performance}

Another avenue of research is using multi-cores to model multi-cores. The USL
model is an excellent candidate for running thousands of simultaneous
regressions in parallel and then selecting an optimal set of coefficients from
the simulation results. Moreover, since the foundations of the USL lie in
queue-theoretic models this approach could be extended to Monte Carlo
simulations~\cite{mcR} of Markov models. \index{multi-core} \index{Monte Carlo}
\index{Markov process}

\section{Concluding Remarks} \label{sec:summary}
In this chapter, we have presented a performance methodology for quantifying
application scalability on multi-core and many-core systems. With potentially
massive computational horsepower now being delivered in low-cost silicon black
boxes, the remaining opportunities for improving performance lie mostly in the
application layers. \index{scalability} \index{methodology} \index{multi-core}

Our methodology, based on the universal scalability law (USL),
emphasizes the importance of validating scalability
data through controlled measurements that use appropriately
designed test workloads. These measurements must then be reconciled with the 
USL performance model. It is this synergy between measuring and modeling that 
provides the key to achieving successful scalability on multi-core platforms.
\index{scalability} \index{Universal scalability law (USL)} \index{modeling} 
\index{workload definition} \index{performance}

In Sect.~\ref{sec:uslmodel} we presented the USL model and showed how it can be
combined with nonlinear statistical regression to analyze controlled performance
measurements. In this way, we are able to truly quantify scalability and thereby
assess the cost-benefit of multithreaded applications running on multi-core or
many-core architectures. The USL methodology also provides data validation as a
side-effect of preparing for the more sophisticated regression analysis.
\index{scalability} \index{methodology} \index{multi-core}

In Sect.~\ref{sec:gpus} we presented some initial results from quantifying GPU
and many-core scalability using the USL methodology. Possible confounding
effects between the USL coefficients due to these fine-grained parallel workloads
suggests an analysis based on the concept of {\em scalability
zones}~\cite{zones}. \index{scalability}

With ever-increasing  economies of scale offered by commodity
multi-core to many-core systems, we anticipate that cost-benefit analysis tools, such as
the USL-based methodology described here, to play an increasingly important role
in the future of computing. \index{methodology}

\bibliographystyle{unsrt} 
\chapbblname{wileychap}
\chapbibliography{wileychap}

\end{document}